\documentclass[aps,twocolumn,prd,aps,floatfix,showpacs,showkeys,superscriptaddress,nofootinbib]{revtex4-2}

\usepackage{amssymb}
\usepackage{amsbsy}
\usepackage{amsthm}
\usepackage{amsmath}
\usepackage{breqn}
\usepackage{bm}
\usepackage{braket}
\usepackage{color}
\usepackage{dcolumn}
\usepackage{epsfig}
\usepackage[T1]{fontenc}
\usepackage{graphicx}
\usepackage[colorlinks]{hyperref}
\usepackage{mathtools}
\usepackage{placeins}
\usepackage{xcolor}
\usepackage{cleveref}
\begin{document}
	
\title{Massive Spatial Qubits: Testing Macro-Nonclassicality \& Casimir Entanglement}
	
	\newcommand{\affone}{Department of Physics and Astronomy, University College London, Gower Street, WC1E 6BT London, United Kingdom.}
\newcommand{\afftwo}{Raman Research Institute, C. V. Raman Avenue, Sadashivanagar, Bengaluru, Karnataka 560080, India}
\newcommand{\affthree} {Center for Astroparticle Physics and Space Science (CAPSS), Bose Institute, Kolkata 700 091, India}
\newcommand{\afffour} {University of Groningen, PO Box 72, 9700 Groningen, Netherlands}
\newcommand{\afffive} {Van Swinderen Institute, University of Groningen, 9747 AG Groningen, Netherlands}

		\author{Bin Yi}
	\affiliation{\affone}

\author{Urbasi Sinha}
	\affiliation{\afftwo}

\author{Dipankar Home}
	\affiliation{\affthree}

\author{Anupam Mazumdar}
	\affiliation{\afffour}
	\affiliation{\afffive}
	\author{Sougato Bose}
	\affiliation{\affone}

	\date{\today}
	
	\begin{abstract}
An open challenge in physics is to expand the frontiers of the validity of quantum mechanics by evidencing nonclassicality of the center of mass state of a macroscopic object. Yet another equally important task is to evidence the essential nonclassicality of the interactions which act between macroscopic objects. Here we introduce a new tool to meet these challenges: massive spatial qubits.  In particular, we show that if two distinct localized states of a mass are used as the $|0\rangle$ and $|1\rangle$ states of a qubit, then we can measure this encoded spatial qubit with a high fidelity in the $\sigma_x, \sigma_y$ and $\sigma_z$ bases simply by measuring its position after different duration of free evolution. This technique can be used reveal the irreducible nonclassicality of the spin-centre of mass entangled state of a nano-crystal implying macro-contextuality. Further, in the context of Casimir interaction, this offers a powerful method to create and certify non-Gaussian entanglement between two neutral nano-objects. The entanglement such produced provides an empirical demonstration of the Casimir interaction being inherently quantum.

		
	\end{abstract}
	
	\maketitle
\paragraph*{\label{sec:intro}Introduction.} It is an open challenge to witness a nonclassicality in the behavior of the center of mass of a massive object \cite{leggett2002testing,arndt2005probing}.  While there are ideas to observe nonclassicalities of ever  more massive objects \cite{bose1999scheme,armour2002entanglement,marshall2003towards,bose2006qubit,chang2010cavity,romero2010toward,romero2011large,romero2011quantum,scala2013matter,wan2016free,bose2018nonclassicality}, the state of art demonstrations  have only reached up to macro-molecules of $10^4$ amu mass \cite{gerlich2011quantum,fein2019quantum}.  Such demonstrations would test the limits of quantum mechanics \cite{diosi1989models, penrose1996gravity, bassi2013models, milburn1991intrinsic, oppenheim2018post, PhysRevA.85.062116},  would be a stepping stone to witness the quantum character of gravity \cite{bose2017spin, marshman2020locality, marletto2017gravitationally, margalit2020realization, van2020quantum}, and  would open up unprecedented  sensing opportunities \cite{marshman2020mesoscopic}.  Identifying new tools to probe  macroscopic nonclassicality (by which here we mean in terms of large mass) is thus particularly important.   Here we propose and examine the efficacy of precisely such a tool: a mechanism to read out a qubit encoded in the spatial degree of freedom of a {\em free (untrapped) mass} (a purely spatial qubit). A principal merit of this scheme is that {\em measuring} the spatial qubit operators $\sigma_x, \sigma_y$ and $\sigma_z$ exploits solely the free time evolution of the mass (Hamiltonian $H=\hat{p}^2/2m$), followed by the detection of its position. As the mass is not controlled/trapped by any fields during its free evolution,  decoherence is minimized.  

 As a first application, we show that our tool enables the verification of an irreducible nonclassicality of a particular joint state of a spin (a well established quantum degree spin) and the center of mass of a macroscopic object, whose quantum nature is yet to be established. To this end, we use the state produced in a Stern-Gerlach apparatus which is usually written down as an {\em entangled} state of a spin and the position \cite{englert1988is,home2007aspects,keil2020stern,margalit2020realization,margalit2019analysis,rosenfeld2021efficient}. Such Stern-Gerlach states have been created with atoms with its spatial coherence verified after selecting a specific spin state \cite{Folman2013,margalit2019analysis}. However, there are, as yet, no protocols to verify the {\em entanglement} between the spin of an object in a Stern-Gerlach experiment and the motion of its center of mass in a way which can be scaled to macroscopic objects. We show that this can be accomplished via the violation of a Bell's inequality in which the spin and the positions of the mass are measured. This violation will also prove the nonclassicality of a large mass in terms of quantum contextuality \cite{basu2001bell,home1984bell}.  

 Next, we propose a second application once the quantum nature of the center of mass degree of freedom of macroscopic objects is assumed (or established in the above, or in some other way). This application has import in establishing the quantum nature of the {\em interactions} between macroscopic objects. We show how our spatial qubit methodology can enable witnessing the entanglement created between two neutral nano-crystals through their Casimir interaction. This has two implications: (a) It will empirically show that the extensively measured Casimir interaction \cite{lamoreaux2012casimir,xu2017detecting,xu2021casimir} is indeed quantum (e.g., is mediated by virtual photons similar to \cite{osnaghi2001coherent,landig2019virtual} -- if photons are replaced by classical entities they would not entangle the masses \cite{bose2017spin,marshman2020locality,marletto2017gravitationally,kafri2013noise,krisnanda2017revealing}). (b) As the entangled state is generated by starting from a superposition of localized states, it is non-Gaussian. While there are ample methods for generating\cite{qvarfort2020mesoscopic, krisnanda2020observable, weiss2020large} and testing \cite{simon2000peres} Gaussian entanglement of nanocrystals, there is hardly any work on their non-Gaussian counterparts. 

We are achieving our tool by combining ideas from two different quantum technologies: photonic quantum information processing and the trapping and cooling of nano-crystals. In the former field a qubit can be encoded in the spatial mode of a single photon by passing it through an effective Young's double slit \cite{taguchi2008measurement}. These qubits, called Young qubits, and their d-level counterparts \cite{ghosh2018spatially,sawant2014nonclassical}, have been exploited in quantum information  \cite{kolenderski2012aharon,rengaraj2018measuring}.  On the other hand, we have had a rapid development recently in the field of levitated quantum nano-objects  \cite{chang2010cavity,romero2010toward,barker2010cavity} culminating in their ground state cooling and the verification of energy quantization \cite{delic2020cooling,tebbenjohanns2019cold}.  While several schemes for verifying quantum superposition of distinct states of such objects have been proposed to date, in these schemes, either the $x, y$ and $z$ motions are measured as infinite dimensional systems \cite{romero2011quantum, bateman2014near, yin2013large} rather than being discretized as an effective qubit, or never measured at all (only ancillary systems coupled to them are measured \cite{scala2013matter, wan2016free}).  Here we adapt the idea of Young qubits from photonic technologies to massive systems. Note that a very different encoding of a qubit in the continuous variables of a harmonic oscillator was proposed long ago for quantum error correction \cite{gottesman2001encoding},  which is not suited to an untrapped nano-crystal.
\begin{figure}
	\includegraphics[width=1
	\columnwidth]{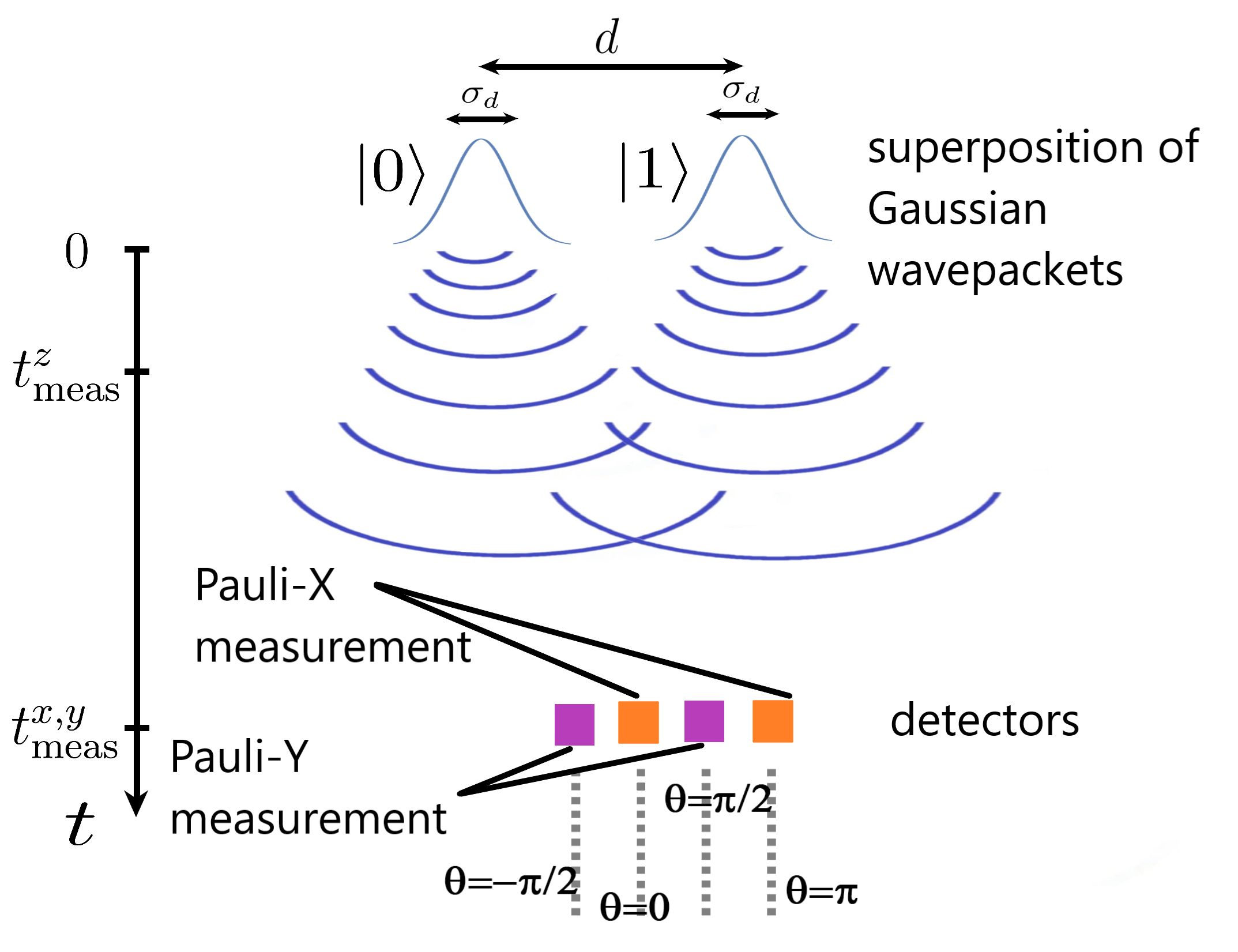}
	\caption{\footnotesize {\bf Spatial detection for $\sigma_x, \sigma_y$ measurements:} a pair of detectors (color: orange) located at phase angle $\theta=0,\theta=\pi$ perform $\sigma_x$ measurement. The detectors (color: purple) at $\theta=\pi/2,\theta=-\pi/2$ perform $\sigma_y$ measurement.} \label{fig1} 
\end{figure}

\paragraph*{\label{sec:stprepr} Qubit Encoding and its Measurement in all Bases:}
Our encoding is intuitive: $|0\rangle$ and $|1\rangle$ states of a qubit are represented by two spatially separated (say, in the $x$ direction) non-overlapping wavepackets whose position and momenta are both centered around zero in the other two commuting ($y$ and $z$) directions. Explicitly, these states (writing only the $x$ part of their wavefunction) are
\begin{equation}
|0\rangle = \frac{1}{\sqrt{\sigma_d} \pi^{1/4}}\int_{-\infty}^{\infty} \exp[-\frac{(x+d/2)^2}{4\sigma_d^2}] |x\rangle dx 
\end{equation}
\begin{equation}
|1\rangle = \frac{1}{\sqrt{\sigma_d} \pi^{1/4}}\int_{-\infty}^{\infty} \exp[-\frac{(x-d/2)^2}{4\sigma_d^2}] |x\rangle dx 
\end{equation}
with $d>>\sigma_d$. These states are schematically depicted in Fig.\ref{fig1} in which only the $x$ direction is depicted along with their evolution in {\em time}. 
For simplicity,  we will omit the acceleration due to the Earth's gravity (as if the experiment is taking place in a freely falling frame),  which can easily be incorporated as its effect commutes with the rest. Thus we only consider 1D time evolution in the $x$ direction.  In this paper, we will only require two states: (a) a state in which a spin embedded in a mass is entangled with the mass's spatial degree of freedom in the state $|\phi^+\rangle=\frac{1}{\sqrt{2}}(|\uparrow,1\rangle+|\downarrow,0\rangle)$ for our first application, and (b) the spatial qubit state $|+\rangle=\frac{1}{\sqrt{2}}(|0\rangle+|1\rangle)$ as a resource for our second application.  Preparation of the above adapts previous proposals and will be discussed with the respective applications. 

We now outline our central tool: the method of measuring the above encoded spatial qubit in various bases. The spatial detection can be performed by shining laser light onto the test masses \cite{PhysRevLett.109.103603,vovrosh2017parametric}. The Rayleigh scattered light field acquires a position dependent phase shift. The scheme is limited only by the standard quantum limit\cite{PhysRevLett.109.103603} (quantum back action) of phase measurement when a {\em large number of photons} are scattered from the mass.  The resolution scales with the number $n$ of scattered and detected photons as $\lambda/\sqrt{n}$,  hence, the power collected at the detector (see Eq.(13) of Ref\cite{vovrosh2017parametric}), and the detection time (as long as this is lower than the dynamical time scale, it is independent of whether the particle is trapped/untrapped).  Thus the detection time should be as much as one needs for the required resolution,  but much less than the time span of the experiment.  By the above methodology,  for a 60nm diameter silica particle, the detection resolution can reach $200\pm20$ fm/$\sqrt{\text{Hz}}$ with laser power $\sim 385\mu$W at the detector, at environmental pressure $\sim0.01$mbar \cite{vovrosh2017parametric}. Thus for an integration time of $\sim4\mu s$, the resolution reaches $\sim1$\AA, which corresponds to just $\sim 10^8$ photons.  As the whole measurement is $\mu s$, any noise of frequency lower than MHz will not affect it (simply remains constant during each measurement run). Moreover, lower frequency noise causing variation between, say, groups of runs, could be measured efficiently by other proximal sensors and taken into account. Also note that the spatial detection is performed at the very end of the protocol, so the question of back action on further position measurements does not arise. 

Due to the spreading of the wavepackets along $y$ and $z$ directions,  when we determine whether the object is in a given position $x=x_0$ at some measurement time $t$, we are essentially integrating the probability of detecting it over a finite region $\Delta y (t)$ and $\Delta z(t)$.  The operator $\sigma_z=|0\rangle\langle0|-|1\rangle\langle1|$ is trivial to measure, as we simply shine a laser centered at $x=d/2$ much before the wavepacket states $|0\rangle$ and $|1\rangle$ have started to overlap (at a time $t^z_{\text{meas}} << d (2\sigma_dm)/\hbar$; the error in $\sigma_z$ measurement as a function of  $t^z_{\text{meas}}$ is described in the supplementary materials (SM); timing errors $\delta t << t^z_{\text{meas}}$ have very little effect).  As described in the previous paragraph,  if $\sim 10^8$ photons are scattered and collected, we can tell apart two states  $|0\rangle$ and $|1\rangle$ separated by $\sim 1$ \AA.

  To measure the spatial qubit $\sigma_x$ and $\sigma_y$ operators, we need a large enough time $t^{x,y}_{\text{meas}} \geq d (2\sigma_dm)/\hbar$ so that the wavepackets of the $|0\rangle$ and $|1\rangle$ states have spread out enough to significantly overlap with each other and produce an interference pattern.  Moreover, due to the free propagation, we would expect the measurement time $t^{x,y}_{\text{meas}}$, final position $x$ and the transverse wave vector $k_x$ are related by: $x=\frac{\hbar k_xt^{x,y}_{\text{meas}}}{m}$ (detecting at a position $x$ after the interference effectively measures the initial superposition state of $|0\rangle$ and $|1\rangle$ in the $|k_x\rangle$ basis). Noting the momentum representation of the qubit states $|n\rangle=\int \left \{ \sqrt2\sigma_d\exp[\frac{ik_xd}{2}-k_x^2\sigma_d^2]\exp[-ink_xd]|k_x\rangle \right \}dk_x$ ($n=0,1$),
the probability to detect the object at a position $x$ for any initial qubit state $|\psi\rangle$ is given by 
\begin{equation}\label{11}
 P(x)=|\langle\psi|k_x\rangle|^2 \propto \left | \exp[\frac{ik_xd}{2}-k_x^2\sigma_d^2]\langle\theta|\psi\rangle \right |^2
\end{equation}
where $|\theta\rangle=|0\rangle+|1\rangle e^{i\theta}$ in which the parameter $\theta=k_xd=\frac{xmd}{\hbar t^{x,y}_{\text{meas}}}$ (we will call $\theta$ the phase angle). Therefore, finding the object in various positions $x$ is in one to one correspondence with positive operator valued measurements (POVM) on the spatial qubit, with the relevant projection on the state being, up to a normalization factor, as $|\theta\rangle\langle\theta|$.  $\sigma_x$ measurements can therefore be implemented by placing a pair of position detectors (which will, in practice be, lasers scattering from the object) at positions corresponding to phase angle $\theta=0,\theta=\pi$; Similarly,  $\sigma_y$ measurements can be achieved by placing detectors at $\theta=\pi/2,\theta=-\pi/2$ (Schematic shown in Fig\ref{fig1}).  For minimizing the time of the experiment, we are going to choose $t^{x,y}_{\text{meas}} = d (2\sigma_dm)/\hbar$. The efficacy of the $\sigma_x$ and $\sigma_y$ measurements as a function of the finite time $t^{x,y}_{\text{meas}}$ for various ratios $\sigma_d:d$ is discussed in the SM.

\begin{figure}
	\includegraphics[width=1
	\columnwidth]{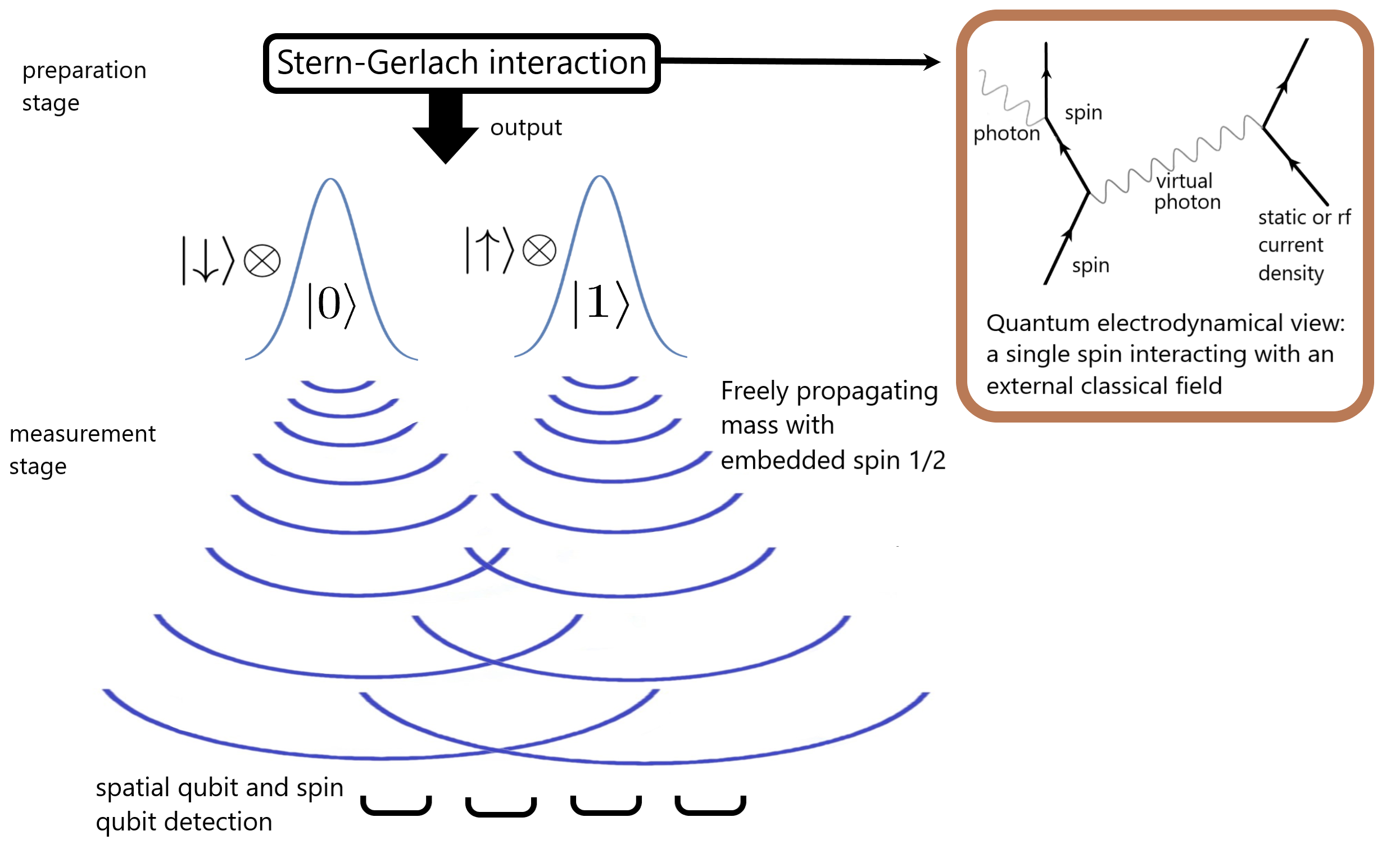}
	\caption{\footnotesize \textbf{Detection scheme for the entanglement of spin and center of mass of a Stern-Gerlach state:} A spin bearing nano-object is measured to be in a set of zones of size $\delta x$, where the size is set by the strength and duration of lasers scattered from the object, which serves to measure the spatial qubit. Within each spatial zone a suitable method is used to measure the spin in different bases, for example by rotating the spin states by microwave pulses followed by fluorescence of certain states under excitation by a laser of appropriate frequency.} \label{fig2} 
\end{figure}

\paragraph*{\label{sec:stprepr} Nonclassicality of the Stern-Gerlach state:}
As a first application of this spatial qubit technology, we consider an extra spin degree of freedom embedded in a mesoscopic mass.  We now imagine that the mass goes through a Stern-Gerlach apparatus. The motion of the mass relative to the source of the inhomogeneous magnetic field (current/magnets) is affected in a spin dependent manner due to the exchange of virtual photons between the source and the spin (Fig.\ref{fig2}) resulting in an entangled state of the spin and position of the nano-object as given by $|\phi^+\rangle=\frac{1}{\sqrt{2}}(|\uparrow,1\rangle+|\downarrow,0\rangle)$, as depicted as the output of the preparation stage in Fig.\ref{fig2}.  It could also be regarded as an intra-particle entanglement (an entanglement between two degrees of freedom of the same object), which has been a subject of several investigations \cite{basu2001bell,PhysRevLett.97.230401,azzini2020single}. 

To measure the spin-motion entanglement in $|\phi^+\rangle$,  we have to measure variables of spin and spatial qubit. Here we specifically want to estimate the action of measuring one of these qubits on the quantum state of the other. During this measurement, the inhomogeneous magnetic field causing the Stern-Gerlach splitting is simply switched off so that spin coherence can be maintained using any dynamical decoupling schemes as required\cite{bar2013solid}. Alternatively, one can also use more pristine nano diamond with less surface defects. As shown in Fig.\ref{fig2}, after a required period of free evolution $t^{x,y}_{\text{meas}}$, measurements of the spatial qubit operators are made; the light shone on the object should not interact at all with the embedded spin degree of freedom if it is completely {\em off-resonant} with any relevant spin transition. Immediately after measuring the spatial qubit, the spin degree of freedom is directly measured in various bases. The latter could be implemented, for example, with a NV center spin qubit in a nano-diamond crystal, where the spin state is rotated by a microwave pulse, which corresponds to basis change, followed by a fluorescence measurement by shining a laser resonant with an optical transition \cite{hensen2015loophole}. The implementation would require cryogenic temperature of the diamond\cite{zhou2014quantum,wang2020coherent}. So the spin coherence time is much greater than the experimental time scale\cite{bar2013solid}. As the spin measurement is very efficient, we only need to consider the resolutions $\delta x, \delta t$ of the spatial qubit measurements so that the effective spatial Pauli $X$ and $Y$ operators are then projections onto a mixed state with phase angle ranging from $\theta-\frac{\delta\theta}{2}$ to $\theta+\frac{\delta\theta}{2}$ with 
$\delta\theta=\frac{ md}{\hbar t^{x,y}_{\text{meas}}}\delta x - \frac{xmd }{\hbar (t^{x,y}_{\text{meas}})^2}\delta t$. For purposes of coherence, which continuously decreases with time, it is best to choose time of the order of the minimum allowed time for overlap of the wavepackets, i.e., choose $t^{x,y}_{\text{meas}} = d (2\sigma_dm)/\hbar$ so that $\delta\theta=\frac{\delta x}{2\sigma_d} - \frac{x \hbar}{4\sigma_d^2 m d}\delta t$. The approximate Pauli matrices are then:
\begin{eqnarray}\label{12,}
\tilde{\sigma}_x&=&
\frac{1}{\delta\theta}
\begin{pmatrix}
0 & \int_{\theta-\frac{\delta\theta}{2}}^{\theta+\frac{\delta\theta}{2}}e^{-i\theta}d\theta|_{\theta=0}\\ \int_{\theta-\frac{\delta\theta}{2}}^{\theta+\frac{\delta\theta}{2}}e^{i\theta}d\theta|_{\theta=0}
& 0
\end{pmatrix} \nonumber
\\\nonumber
&=&\frac{1}{\delta\theta}
\begin{pmatrix}
0 & -ie^{i\frac{\delta \theta}{2}}+ie^{-i\frac{\delta \theta}{2}}\\ -ie^{i\frac{\delta \theta}{2}}+ie^{-i\frac{\delta \theta}{2}}
& 0
\end{pmatrix}
\nonumber
\end{eqnarray}
and similarly,
	$\tilde{\sigma}_y=\frac{1}{\delta\theta}
	\begin{pmatrix}
		0 & -e^{i\frac{\delta \theta}{2}}+e^{-i\frac{\delta \theta}{2}}\\ e^{i\frac{\delta \theta}{2}}-e^{-i\frac{\delta \theta}{2}}
		& 0
	\end{pmatrix}.$
To verify the entanglement we have to show that the spin-motion entangled state violates the Bell-CHSH inequality $
{\cal B} = |\langle AB\rangle+\langle A{B}'\rangle+\langle {A}'B\rangle-\langle {A}'{B}'\rangle|\leq 2$
with variables\cite{PhysRevLett.24.549}  $A=\tau_x + \tau_y$ and $A'=\tau_x-\tau_y$ operators of the spin ($\tau_x$ and $\tau_y$ are spin Pauli matrices) and $B=\tilde{\sigma}_x$ and $B'=\tilde{\sigma}_y$ operators of the spatial qubit.  The expected correlation can be calculated (see SM) to give ${\cal B} = |2\sqrt2 f(\delta \theta)|\leq2\sqrt2$
where $f(\delta\theta)=\frac{2}{\delta\theta}Re[ie^{i\delta\theta/2}]=\frac{2}{\delta\theta}cos(\frac{\pi+\delta\theta}{2})$.  To obtain a violation of the CHSH inequality the upper bound of $\delta\theta$ can be calculated:$|f(\delta\theta)|=\frac{1}{\sqrt2},\delta\theta$$\approx2.783$. Due to high tolerance in spatial detection, one may pick $\delta \theta=\pi/2$ so that the four detectors consist of Pauli-X and Y measurements are placed adjacent to each other and cover the full range $\theta\in\left[-3/4\pi,5/4\pi\right]$. The probability of detection is $\sim17.7\%$ for each repetition of the experiment.

 For realization,  consider a $m \sim 10^{-19}$ kg ($10^8$ amu) spin-bearing mass cooled to a ground state in $\omega\sim1$ kHz trap  \cite{zheng2020room,leng2021mechanical} so that its ground state spread is $=\sqrt{\frac{\hbar}{2m\omega}}\sim 1$nm. The cooling to ground state essentially requires a measurement to the accuracy of the ground state spread and sufficient isolation. Essentially, measuring the position of an object to the above precision (nm) requires $3\times10^7$ photons, which will hardly heat up the system in a diamagnetic trap. In fact, feedback cooling has been achieved for massive ($10$kg) masses \cite{whittle2021approaching}.     
 
 At time $t=0$ the embedded spin is placed in a superposition $1/\sqrt{2}(|\uparrow\rangle+|\downarrow\rangle)$, and the mass is released from the trap. The wave packet then passes through an inhomogeneous magnetic field gradient $\sim 10^5$ Tm$^{-1}$ \cite{margalit2020realization}. Due to the Stern-Gerlach effect, the mass moves in opposite directions corresponding to $|\uparrow\rangle$ and $|\downarrow\rangle$ spin states, and, in a time-scale of $t_{\text{prep}} \sim 50 \mu$s, evolves to a $|\phi^{+}\rangle$ state with a separation of $d=25$nm  between the $|0\rangle$ and $|1\rangle$ spatial qubit states \cite{wan2016free, bose2017spin, marshman2020mesoscopic, van2020quantum, margalit2020realization} (all lower $m$ and $d$ are also possible as they demand lower $t_{\text{prep}}$ and $\partial B/\partial x$). To keep the spin coherence for $t_{\text{prep}}$, dynamical decoupling may be needed \cite{bar2013solid}; it is possible to accomodate this within our protocols-one just needs to change the direction of the magnetic field as well in tandem with the dynamical decoupling pulses which flip the spin direction\cite{wood2022spin}. During the above $t_{\text{prep}}$, the spread of wavepackets is negligible so that $\sigma_d$ remains $\sim 1$nm. 
 
According to our results above, in order to obtain a CHSH inequality violation, one has to measure to within $\delta x \sim 2 \sigma_d \delta \theta \sim 1$nm  resolution. To achieve this resolution, firstly, we have to ensure that during the whole duration of our protocol, the acceleration noise has to below a certain threshold so as to not cause random displacement greater than 1nm. Given $t^{x,y}\sim50$ms is the longest duration step, the acceleration noise needs to be $\sim10^{-6}$ms$^{-2}$. Next comes the measurement step where light is scattered from the object, which also needs to measure to the required resolution. This is possible as there are feasible techniques that give resolutions of 0.1 pm/$\sqrt{\text{Hz}}$ \cite{PhysRevLett.109.103603,vovrosh2017parametric} for position measurements by scattering light continuously from an object. Adopting the scheme in \cite{vovrosh2017parametric}, the resolution can be achieved by scattering light continuously from the object for about $1\mu$s, which is $4$ orders of magnitude smaller than the experimental time span. On the other hand if the timing accuracy $\delta t$ of $t^{x,y}_{\text{meas}}$  is kept below $\sim 0.1$ms (also easy in terms of laser switching on/off times), there is a negligible inaccuracy in $\theta$.

  Note that as shown in the SM, dephasing between the spatial states $|0\rangle$ and $|1\rangle$ at a rate $\gamma$ simply suppresses the CHSH violation by a factor $e^{-\gamma t}$, which could be a new way to investigate decoherence of the mass from various postulated models \cite{diosi1989models, penrose1996gravity, bassi2013models, milburn1991intrinsic, oppenheim2018post, PhysRevA.85.062116} and environment. 
  
 The decoherence of the spatial degree of freedom results from background gas collision and black-body radiation.  Adopting the formulae from Ref\cite{romero2011quantum},  for our realization, the contribution to $\gamma$ from background gas reaches $\sim167.2$ s$^{-1}$ at pressure $\sim10^{-10}$Torr.  Black-body radiation induces decoherence at a rate of $\sim274.9$s$^{-1}$ at internal temperature $50$K.  On the other hand, the spin degree of freedom, which may be encoded in NV centers of nano-diamond crystal, reaches a coherence time of $\sim0.6$s at liquid nitrogen temperature $77$K\cite{bar2013solid}. As it stands, the coherence of electron spins in nano-diamonds is lower than what we require for $10^{-19}$kg mass \cite{naydenov2011dynamical,tisler2009fluorescence,laraoui2012nitrogen}. However, recently, much larger times of $0.4$ms has been achieved via dynamical decoupling \cite{wood2022long}. This is already much larger than the superposition state preparation time ($\sim50\mu$s). There are ideas to incorporate dynamical decoupling in the preparation of the initial superposition in our experiment \cite{wood2022spin}. There are also ideas of using bath dynamical decoupling which should not affect the spatial superposition generation \cite{knowles2014observing}. Note that, strictly speaking, the electronic spin is only required during the Stern-Gerlach generation of superposition after which it can be mapped onto nuclear spins. Before the free packet expansion (interferometry), the electronic spin can be mapped onto nuclear spin which has much longer coherence times. At the end of the protocol, we will need to measure the nuclear spin, for which we may need to map back to electronic state. Such mapping can happen in micro-second given the hyperfine couplings \cite{wood2021quantum}. Therefore, achievable pressure and temperature make the coherence time sufficient for the realization of our protocol (same estimates hold for the protocol of the next section). 
  
  As {\em both} the  spin and the mass are measured, it characterizes the entanglement of the given state {\em irrespective} of the dynamics from which the state was generated, as opposed to previous protocols which rely on a reversible nature of the quantum dynamics  \cite{bose1999scheme,armour2002entanglement,marshall2003towards,bose2006qubit}. As opposed to single object interferometry \cite{romero2011quantum, bateman2014near, yin2013large},  here the CHSH violation explores decoherence of the mass in multiple bases -- not only how the $|0\rangle \langle 1|$ term of the spatial qubit decays (position basis),  but also whether, and, if so, how, $|+\rangle \langle -|$ decays (where $|-\rangle = |0\rangle - |1\rangle$) -- a novel type of decoherence of even/odd parity basis. Moreover, as the total spin-motional system is quantum 4 state system, the CHSH violation can also be regarded as a violation of the classical notion of non-contextuality \cite{home1984bell,basu2001bell,cabello2008experimentally}. 

\begin{figure}
	\includegraphics[width=1
	\columnwidth]{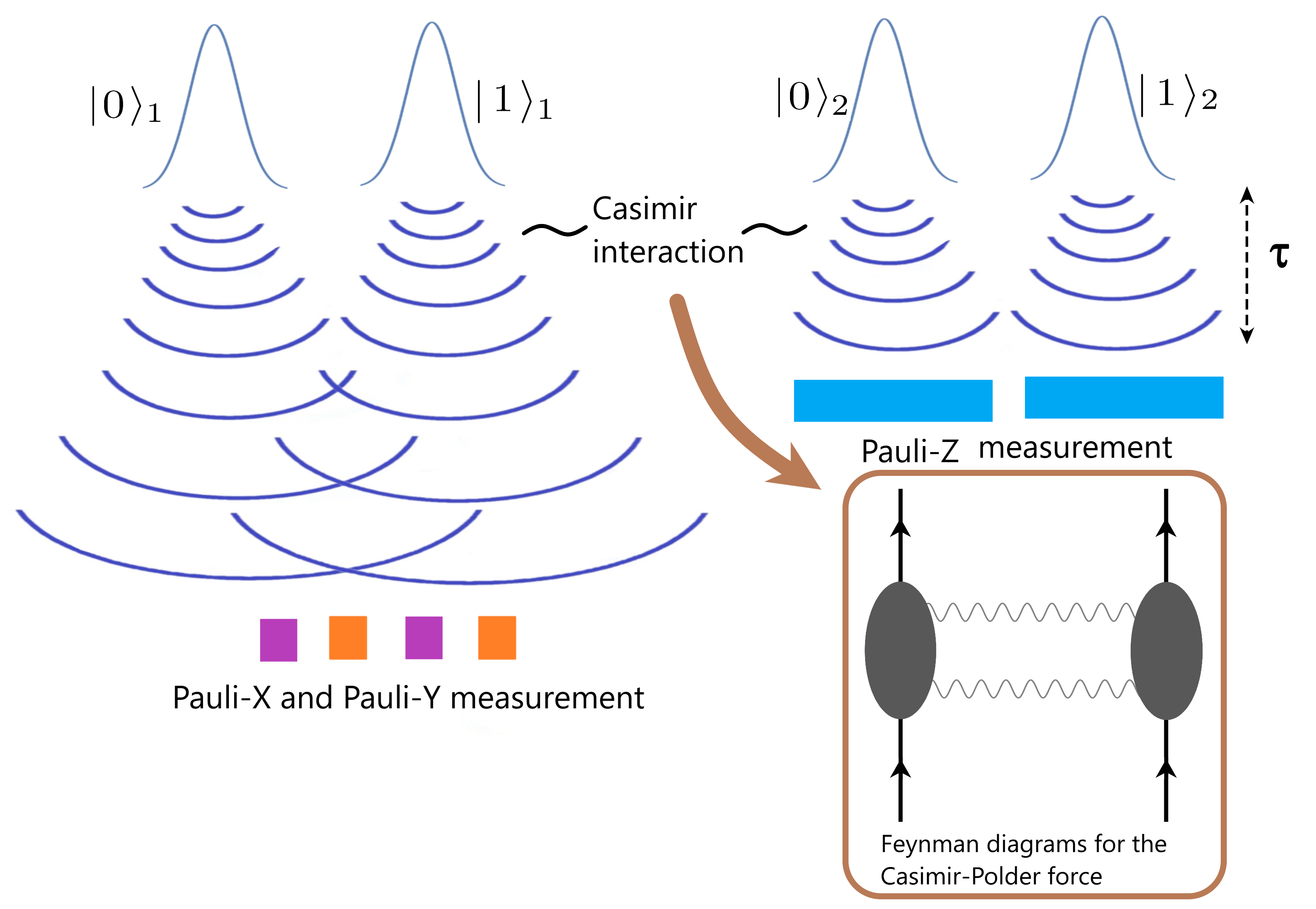}
	\caption{\footnotesize \textbf{Application in witnessing Casimir induced entanglement:} Two masses, each prepared in a superposition of two states, act as two qubits $\frac{1}{\sqrt2}(|0\rangle_1+|1\rangle_1)\otimes\frac{1}{\sqrt2}(|0\rangle_2+|1\rangle_2)$. The system freely propagates and undergoes mutual interactions for a time $\tau$. This interaction induces entanglement which can be witnessed from correlations of spatial qubit Pauli measurements.  For example, in the figure,  $\sigma_x$, $\sigma_y$ measurements on test mass $1$ and $\sigma_z$ measurements on test mass $2$ are depicted. Casimir interaction induced by virtual photons as quantum mediators is shown \cite{feinberg1970general}.} \label{fig3} 
\end{figure}

\paragraph*{\label{sec:stprepr} Casimir induced  entanglement:}
Neutral unmagnetized untrapped masses, ideal for the preservation of spatial coherence, can  interact with each other via the Casimir interaction \cite{van2020quantum} (gravity can cause observable effects in reasonable times only for masses $>10^{-15}-10^{-14}$ kg \cite{bose2017spin,van2020quantum}).  Two such masses (mass $m$, radius $R$) indexed 1 and 2 are each prepared in the spatial qubit state $|+\rangle$ (the  superposition size, separation between states $|0\rangle$ and $|1\rangle$,  being $d$) while the distance between the centers of the superpositions is $D$ (Fig.\ref{fig3}). In a time $\tau$, the Casimir interaction evolves the system to
\begin{eqnarray}
\frac{e^{i\phi}}{\sqrt2} [|0\rangle_1\frac{1}{\sqrt2}(|0\rangle_2+e^{i\Delta \phi_{01}}|1\rangle_2)\nonumber\\
+|1\rangle_1\frac{1}{\sqrt2}(e^{i\Delta \phi_{10}}|0\rangle_2+|1\rangle_2)]
\end{eqnarray}
where $\phi=k\frac{R^6}{D^7}\tau, \Delta\phi_{01}=k\frac{R^6}{(D+d)^7}\tau-\phi, \Delta\phi_{10}=k\frac{R^6}{(D-d)^7}\tau-\phi$,  in which
$k=\frac{23c}{4\pi}(\epsilon-1)^2/(\epsilon+2)^2$\cite{casimir1948influence} (See also \cite{pedernales2020motional,kim2005static}), where $\epsilon$ is the dielectric constant of the material of the masses. This formula is valid when the separation is much larger than the radius of the sphere, which is the regime of our subsequent calculations. On top of the above evolution, we assume a local dephasing  ($|0\rangle \langle 1|_i \rightarrow e^{-\gamma \tau} |0\rangle \langle 1|_i$) for both particles ($i=1,2$) (this can generically model all dephasing \cite{van2020quantum, torovs2020relative}). 
To verify the induced entanglement, one can make spatial qubit measurements up to uncertainties parametrized by $\delta \theta$ as outlined previously and then estimate the entanglement witness \cite{PhysRevA.102.022428} $W=I\otimes I-\tilde{\sigma}_x \otimes \tilde{\sigma}_x-\tilde{\sigma}_z\otimes \tilde{\sigma}_y- \tilde{\sigma}_y\otimes \tilde{\sigma}_z$ where $\tilde{\sigma}_x$ and $\tilde{\sigma}_y$ are as discussed before, and we take $\tilde{\sigma}_z = \int_{-\infty}^0 |x\rangle\langle x| dx -  \int_{0}^{\infty} |x\rangle\langle x| dx$. If $ \langle W\rangle=Tr(W\rho)$ is negative, the masses are entangled.  We find
\begin{align}
\langle \tilde{W} \rangle=&1-\frac{1}{2}e^{-2\gamma t}g^{2}(\delta\theta)(1+\cos(\Delta\phi_{10}-\Delta\phi_{01}))
\nonumber\\-&e^{-\gamma t}g(\delta\theta)(\sin(\Delta\phi_{10})+\sin(\Delta\phi_{01})),
\label{avg W'}
\end{align}
where $g(\delta\theta)=\frac{2}{\delta\theta}\cos(\frac{\pi-\delta\theta}{2})$.  
  
 We are going to consider the Stern-Gerlach mechanism to first prepare the state $|\phi^+\rangle$, and use that to prepare $|+\rangle$. We consider a $R\sim 20$ nm, $m\sim1.17\times10^{-19}$ kg mass, and consider it to have been trapped and cooled it to its ground state ($\sigma_d \sim 1$ nm) in a 1 kHz trap \cite{yin2013large}. We then release it, and subject it to a magnetic field gradient of $5\times 10^4$ Tm$^{-1}$ \cite{margalit2020realization} for $t \sim 100 \mu$s so that a Stern-Gerlach splitting of $d\approx 50$ nm develops while there is insignificant wavepacket spreading ($\sigma_d$ remains $\sim 1$ nm). At this stage,  a microwave pulse may be given to rotate the spin state so that the $|\phi^+\rangle$ state evolves to  $|0\rangle (|\uparrow\rangle+|\downarrow\rangle) + |1\rangle (|\uparrow\rangle-|\downarrow\rangle)$. A subselection of the $|\uparrow\rangle$ spin state via deflection through another Stern-Gerlach,  then yields the state $|+\rangle$ \cite{Folman2013,margalit2019analysis}.  Alternatively, by performing a Controlled-NOT with the spatial qubit as the control and the spin as the target (again, performed quite accurately by a microwave pulse \cite{yin2013large}), $|\phi^+\rangle$ gets converted to $|+\rangle |\downarrow\rangle$ so that the spatial part is our required state.  For $D\approx2.1 \mu$m, then $\Delta \phi_{10}=\phi_{10}-\phi\approx 0.17$, $\Delta \phi_{01}=\phi_{01}-\phi\approx -0.14$ after $\tau \sim 0.012$s of entangling time, which gives a negative witness $\langle W\rangle \sim-0.0064$. A value of $\delta \theta\sim \pi/6$ is the highest tolerance of error in spatial detection to keep the entanglement witness negative. The position detectors that consist of Pauli-X and Y measurements with width $\delta \theta\sim \pi/6$ has $\sim5.9\%$ chance of detection for each repetition of the experiment.

 Note here that the form of  witness operator compels one to measure both the $\tilde{\sigma_x} \otimes \tilde{\sigma_x}$ operator and the other two operators on the {\em same entangled state}.  $\tilde{\sigma}_z$ measurement is also done at $t^z_{\text{meas}}=\tau$. This is about $0.1$ of the overlapping time $\sim\frac{d(2\sigma_d m)}{\hbar}$ so that the fidelity of the $\sigma_z$ measurement is very high (see SM).  We then require $t^{x,y}_{\text{meas}}-t^z_{\text{meas}} << \tau$ so that the extra entanglement generated due to interactions after the $\tilde{\sigma_z}$ measurement and before the $\tilde{\sigma_x}/\tilde{\sigma_y}$ measurements  is negligible. This, in turn, requires us to {\em speed up} the development of spatial overlap between the qubit states due to wavepacket spreading after the $\tilde{\sigma_z}$ measurement, which can be accomplished by squeezing both of the wavepackets in position after the time $t^z_{\text{meas}}$.  After 0.01 s of flight, the wavepacket width $\sigma_d\sim 1$ nm expands to $\sim 10$ nm. Thus we have to squeeze the state by 2 orders of magnitude to $\sim1\times10^{-10}$ m, so that it expands to $\sim100$ nm, where overlapping occurs, in the next $0.001$.s The fidelity of $XY$ measurement here is very high (see SM). The slight delay in $\sigma_{x/y}$ measurement ($0.001$s later than the $\sigma_z$ measurement) would cause only a $\sim5\%$ error in the witness magnitude.  Note that in order to achieve the required squeezing, two appropriate periods of unitary evolution in harmonic potentials of $\omega_1 \sim$1 MHz and $\omega_2 \sim$ 0.1 MHz would suffice ($n$ repeated changes between $\omega_1$ and $\omega_2$ separated by appropriate periods of harmonic evolution  will squeeze by the factor $(\omega_1/\omega_2)^n$ \cite{KISS1994311}); if this potential was applied as an optical tweezer then it will hardly cause any decoherence $\gamma_{\text{squeeze},j} \sim \omega_j 10^{-5}$ \cite{chang2010cavity}. We additionally need to ensure, for reasons described earlier, that the acceleration noise is below $10^{-6}$ms$^{-2}$. The whole procedure described above could be one of the earliest demonstrations of non-Gaussian entanglement between neutral masses.  It would also demonstrate the nonclassical nature of the Casimir interaction, namely that it is mediated by quantum agents (virtual photons) as in the inset of Fig.\ref{fig3}.
 
 Compare with other types of interactions, gravitational interaction at this scale is negligible compare with Casimir interaction as long as the separation between the two masses are less than $200\mu$m for materials with density of diamonds \cite{bose2017spin}. The phase induced by electrostatic interaction can be estimated by $\frac{\vec{p}^2t}{4\pi\epsilon_0R^3\hbar}$, where $\vec{p}$ is electric dipole moment. For $R\sim20$nm, $t\sim10$ms, $\vec{p}$ needs to be much smaller than $10^-30$C.m to be considered negligible compare with Casimir interaction. In current experiments, $\vec{p}$ can be as small as $\sim10^{-23}$C.m for $10\mu$m radius particles, scaling with the volume of the particle \cite{afek2021control,rider2016search}. From the scaling, we can expect electric dipole $\sim10^{-29}$C.m for $10$nm radius particles. An improvement of the electric dipole by a couple of order of magnitude is required to make electrostatic interaction negligible in comparison with Casimir interaction. These electric dipoles typically appear due to defects in crystals such as dislocations. It is possible that given the $10$ nm size, one can form a single crystal \cite{frangeskou2018pure} so that the intrinsic dipole background can be reduced.

\paragraph*{\label{sec:stprepr} Conclusions:}

We have shown how to measure a qubit encoded in a massive object by position detection.
We have shown how this can be applied to: (a) stretch the validity of quantum physics to the center of mass of nano-objects -- demonstrating quantum contextuality, never before tested for macroscopic objects, (b) entangle spatial qubits encoded in two such objects,  extending non-Gaussian quantum technology, (c) prove empirically the quantum coherent nature of the Casimir force. Indeed, in the {\em same} open-minded way that one asks whether quantum mechanics continues to hold for macroscopic masses \cite{leggett2002testing,arndt2005probing}, one can question whether those interactions between such masses which are extensive in nature (grow as volume/area/mass) {\em continue} to be mediated by a quantum natured field so as to be able to entangle the masses.  In comparison with standard approaches for probing the nonclassicality for smaller masses, we avoid a Mach-Zehnder interferometer -- only requiring the preparation of an original spatial superposition. This is advantageous because of the difficulty of realizing beam-splitters for nano-objects (tunneling probability $\propto e^{-\frac{\sqrt{2mV}}{\hbar}\Delta x}$ getting extremely small), and also for avoiding interactions with mirrors and beam splitters which can cause decoherence.(we exploit a two-slit experiment as a beam-splitter \cite{sadana2019double},  see SM). Our methodology can be quantum mechanically simulated with cold atoms, where other methods to encode qubits in motional states have been demonstrated\cite{fluhmann2019encoding}, before they are actually applied to nano-objects. 

\acknowledgements DH and US would like to acknowledge partial support from the DST-ITPAR grant IMT/Italy/ITPAR-IV/QP/2018/G. DH also acknowledges support from the NASI  Senior Scientist fellowship. AM's research is funded by the Netherlands Organization for Science and Research (NWO) grant number 680-91-119. SB would like to acknowledge EPSRC Grant Nos. EP/N031105/1 and  EP/S000267/1.


%

\end{document}